\documentclass[10pt,twocolumn,letterpaper]{article}

\usepackage{cvpr}      

\usepackage{graphicx}
\usepackage{amsmath}
\usepackage{amssymb}
\usepackage{booktabs}
\usepackage{multirow}
\usepackage[accsupp]{axessibility} 

\usepackage[pagebackref,breaklinks,colorlinks]{hyperref}

\usepackage[capitalize]{cleveref}
\crefname{section}{Sec.}{Secs.}
\Crefname{section}{Section}{Sections}
\Crefname{table}{Table}{Tables}
\crefname{table}{Tab.}{Tabs.}

\def\cvprPaperID{3} 
\def\confName{CVPR}
\def\confYear{2022}

\begin{document}

\title{Towards Comprehensive Testing on the Robustness of Cooperative Multi-agent Reinforcement Learning}

\author{Jun Guo$^{1,2}$, Yonghong Chen$^2$, Yihang Hao$^2$, Zixin Yin$^1$, Yin Yu$^3$, Simin Li$^{1*}$\\
$^1$ State Key Lab of Software Development Environment, Beihang University, Beijing, China\\
$^2$ Yangzhou Collaborative Innovation Research Institute CO., LTD \\
$^3$ No. 38 Research Institute of CETC \\
{\tt\small $\{$junguo, lisiminsimon, yzx835$\}$@buaa.edu.cn,} \\ {\tt\small $\{$13514262035@163.com, hyh19951114$\}$@gmail.com, tony\textunderscore yu\textunderscore0210@126.com}}

\maketitle

\begin{abstract}
    While deep neural networks (DNNs) have strengthened the performance of cooperative multi-agent reinforcement learning (c-MARL), the agent policy can be easily perturbed by adversarial examples. Considering the safety critical applications of c-MARL, such as traffic management, power management and unmanned aerial vehicle control, it is crucial to test the robustness of c-MARL algorithm before it was deployed in reality. Existing adversarial attacks for MARL could be used for testing, but is limited to one robustness aspects (e.g., reward, state, action), while c-MARL model could be attacked from any aspect. To overcome the challenge, we propose MARLSafe, the first robustness testing framework for c-MARL algorithms. First, motivated by Markov Decision Process (MDP), MARLSafe consider the robustness of c-MARL algorithms comprehensively from three aspects, namely state robustness, action robustness and reward robustness. Any c-MARL algorithm must simultaneously satisfy these robustness aspects to be considered secure. Second, due to the scarceness of c-MARL attack, we propose c-MARL attacks as robustness testing algorithms from multiple aspects. Experiments on \textit{SMAC} environment reveals that many state-of-the-art c-MARL algorithms are of low robustness in all aspect, pointing out the urgent need to test and enhance robustness of c-MARL algorithms.
\end{abstract}

 
\section{Introduction}
\label{sec:intro}

With the success of deep neural networks (DNNs), tremendous success have been made in cooperative multi-agent reinforcement learning (c-MARL), which powers numerous real-world applications, including traffic management \cite{greguric2020trafficsignal1,chu2019trafficsignal2}, power management \cite{wang2021voltagecontrol1,fang2019voltagecontrol2} and unmanned aerial vehicle control \cite{cui2019UAVcontrol1,cui2019UAVcontrol2}, \emph{etc}. Recently, it has been shown that adversarial examples \cite{gleave2019adversarial,behzadan2017vulnerability,liu2020spatiotemporal,lin2020robustmarl,liang2021parallel, Wei2021TowardsTA} are capable to perturb these safety-critical application with high-confidence, raising a serious concern on the robustness of c-MARL algorithms.

\begin{figure}[t]
  \centering
  \includegraphics[width=0.95\linewidth]{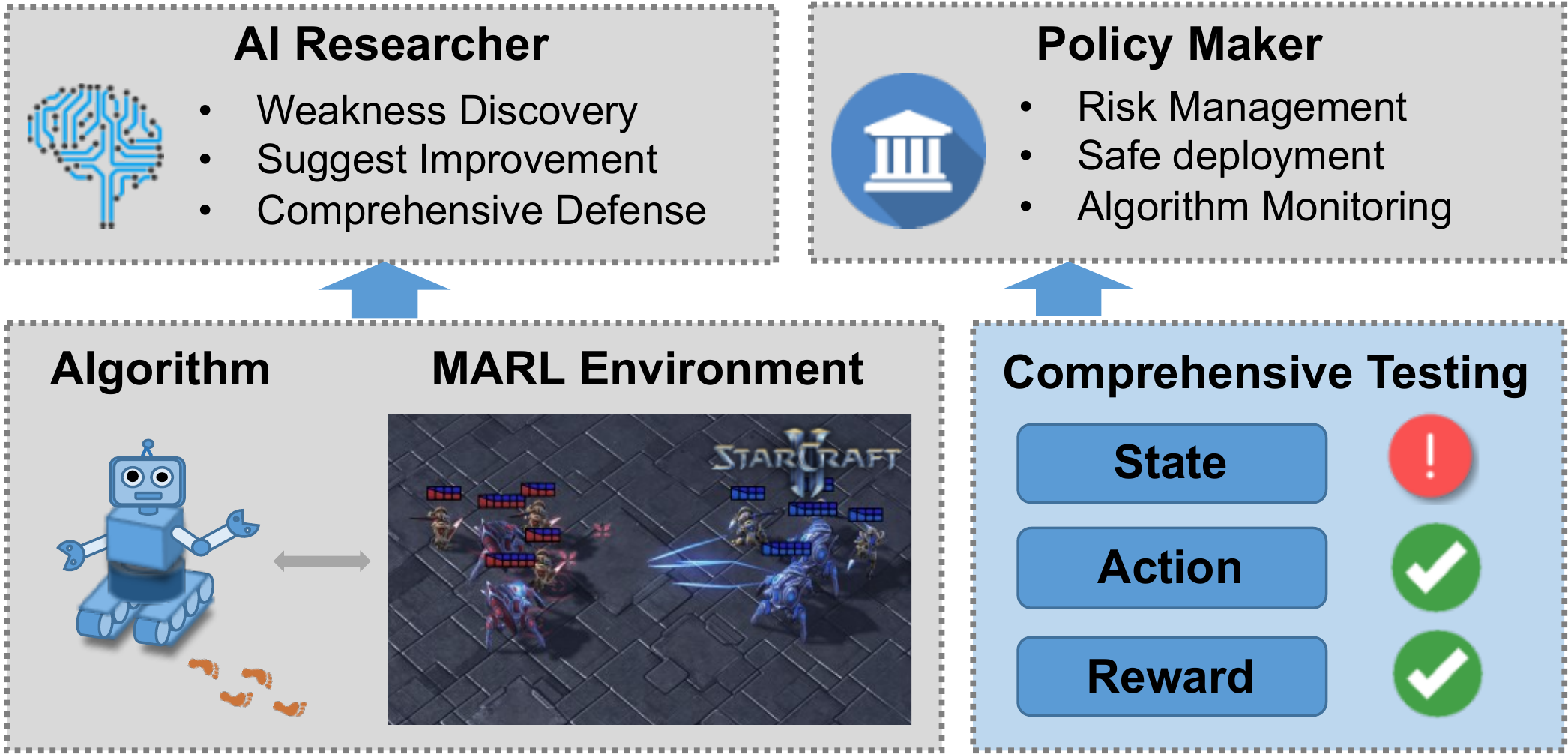}
   \caption{An overview of our work. c-MARL algorithm could be attacked from three aspects, namely state, action and reward. We test the robustness of c-MARL from these aspects.}
   \label{fig:overview}
   \vspace{-0.2in}
\end{figure}

Testing the robustness has been a promising solution for DNN models. Being able to thoroughly test the robustness of DNN models will benefit researchers to discover weakness in DNN models and policy makers to ensure safe deployment in many sensitive scenarios. Many highly-influential works have been published in computer vision communities to test robustness and interpret adversarial examples \cite{gao2020robusttesting1,tang2021robustart,huang2011robusttesting2,kim2019robusttesting3,kurakin2016robusttesting4,liu2021robusttesting5, zhang2020interpreting, liu2021ANP} using multiple algorithms, metrics and attack settings. Recently, Behzadan et al.\cite{behzadan2019rlbenchmark} also benchmarked the robustness of reinforcement learning (RL) algorithms towards different state perturbations.

However, to the best of our knowledge, no work exists to test the robustness of c-MARL algorithm. Besides, from the perspective of multi-agent MDP, its possible for hackers to attack from the aspect of reward \cite{han2018reinforcement}, state \cite{lin2017tactics} and action \cite{gleave2019adversarial}. While existing attack could be used as testing tool, they all focus on only one aspect (state, action, reward), making the test limited since c-MARL algorithm might be robust in one aspect, but hacker can attack from all possible aspects.


To tackle the problem, we propose MARLSafe, a multi-aspect testing framework of c-MARL algorithms. The motivation of our paper is summarized in Fig. \ref{fig:overview}. First, motivated by multi-agent Markov Decision Process (MMDP), multi-agent reinforcement learning contains 5 \textit{elements}: state, action, reward, environment dynamic, and discount factor. Consider existing literature and the feasibility of perturbation, we assume hacker might perturb the state, action and reward in MMDP. Then, we view the robustness of c-MARL as the ability to resilience attacks from multi aspects. Note that since hackers might attack from any aspect, an algorithm must be simultaneously robust in all aspects in order to be considered robust. Second, due to the scarcity in c-MARL attack literature, to the best of our knowledge, \cite{lin2020robustmarl} is the only paper to attack c-MARL in state aspect, while no work exists to attack other aspects. We propose c-MARL attacks based on the aspect of state, action and reward. To figure out the performance and characteristic of these attacks in c-MARL tasks, we conduct experiments of attacks on the \textit{StarCraftII Multi Agent Challenge} (\textit{SMAC}) environment\cite{samvelyan19smac}. Our contributions can be listed as follows:

\begin{itemize}
    
    \item To the best of our knowledge, MARLSafe is the first to test the robustness of c-MARL algorithms from multiple aspects, namely state, action and reward.
    \item Technically, MARLSafe propose adversarial attack for c-MARL algorithms from multiple aspects. Some aspects are first proposed in c-MARL literature.
    \item Empirically, we find the testing method in MARLSafe could attack state-of-the-art c-MARL algorithm with high confidence (\ie, towards 0\% winning rate).
\end{itemize}

\section{Related Work}
\subsection{Adversarial Attacks} Szegedy \etal\cite{szegedy2013intriguing} first defined adversarial attacks and proposed L-BFGS attack to generate adversarial examples. By leveraging the gradient of the target model, Goodfellow \etal\cite{goodfellow2014explaining} proposed the Fast Gradient Sign Method (FGSM) to quickly generate adversarial examples. Since then, many types of adversarial attacks have been proposed, such as gradient-based attacks (PGD, C\&W)\cite{madry2017towards, carlini2017towards}, boundary-based attack (DeepFool)\cite{moosavi2016deepfool}, saliency-based attack (JSMA)\cite{papernot2016limitations}. Brown \etal\cite{brown2017adversarial} first proposed advesarial patch, which adds a local patch with impressive textures to the input image. Liu \etal\cite{liu2020bias} proposed a patch attack towards automatic check-out in physical world. Wang \etal\cite{wang2021dual} proposed Dual Attention Suppression attack to make the adversarial patches both malign and beautiful. Adversarial attacks on machine learning models have been adequately investigated, showing the potential risk of neural networks when it comes to practical application. 

\subsection{Multi-Agent Deep Reinforcement Learning} Deep reinforcement learning methods tend to train a policy network which maps state observations to action probabilities. DRL algorithms can be roughly categorized into two types: policy-based and value-based algorithm. Policy based algorithms often rely on policy gradient, such as DDPG\cite{lillicrap2015continuous} and PPO\cite{schulman2017proximal}. Value based algorithms often predict the Q-value, such as Deep Q Network (DQN)\cite{mnih2013playing}. In MARL tasks, the most straightforward way to acquire a policy is to train individual agents, which is called Independent Q-Learning (IQL)\cite{tan1993multi, tampuu2017multiagent}. However, this strategy is not efficient in MARL environments requiring cooperation. Recent works adopted CTDE framework, such as QMIX\cite{rashid2018qmix} and MAPPO\cite{yu2021surprising}, can enhance the cooperation of agents and achieve better performance. However, those algorithms also suffer from robustness problem, which have not been properly evaluated.

\subsection{Adversarial Attacks on DRL}  Huang \etal \cite{huang2017adversarial} evaluated the robustness of DRL policies by perturbing the observations through FGSM attack on Atari Games. Liu \etal \cite{liu2020spatiotemporal} proposed a spatiotemporal attack for embodied agents, which generates adversarial textures in the navigation environment. Lin \etal \cite{lin2017tactics} proposed an attack method which perturbs the observation at some crucial frames, and they achieved targeted attack for DRL policies. Behzadan and Munir \cite{behzadan2017vulnerability} propose a black-box attack by introducing a surrogate policy to minimize the return. Han \etal \cite{han2018reinforcement} proposed reward flipping attack at train time in software-defined networking tasks. Gleave \etal \cite{gleave2019adversarial} proposed the adversarial policy in competitive multi-agent settings, which trains an opponent agent while fix parameters of the victim policy to attack the victim model. To the best of our knowledge, \cite{lin2020robustmarl} is the only paper to attack c-MARL by perturbing the input state of agent.

\begin{figure*}
  \centering
  \includegraphics[width=0.90\linewidth]{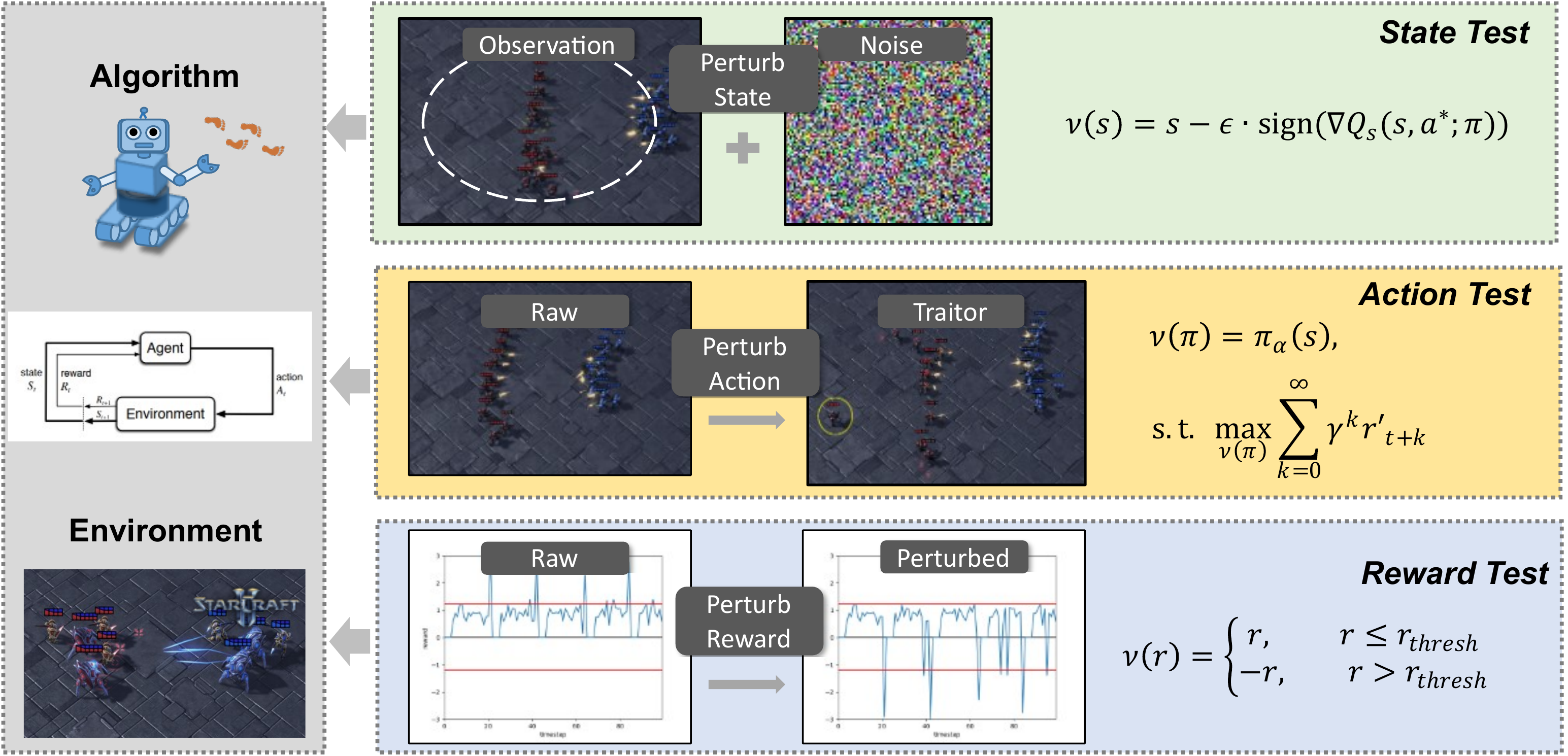}
  \caption{The framework of MARLSafe. Motivated by MMDP framework, we propose state, action and reward test to test its robustness.}
  \label{fig:framework}
  \vspace{-0.2in}
\end{figure*}

\section{Methodology}

\subsection{Formulation of Attacks}

A multi-agent Markov Decision Process (MMDP) is defined as a tuple $(\mathcal{S}, \{\mathcal{A}^i\}_{i \in \mathcal{N}}, R, \mathcal{P}, \gamma)$, where $\mathcal{S}$ denotes the state space, $\mathcal{A}:=\mathcal{A}^1 \times ... \times \mathcal{A}^N $ denotes the joint action space with N agents, $R:\mathcal{S}\times \{\mathcal{A}\} \times \mathcal{S} \rightarrow \mathbb{R}$ is the joint reward function for all c-MARL agents, and $\mathcal{P}: \mathcal{S} \times \{\mathcal{A}\} \rightarrow \mathcal{S}$ is the transition probability of the environment, also known as environment dynamics. The next state is determined by environmental dynamics, current state and actions taken by agent: $\mathcal{P}(s', r|s,\{a^i\})=p(s_{t+1}=s', r_{t+1}=r|s_t=s,\{a_t^i\}=\{a^i\})$, where $t$ is the time step. $\gamma \in [0, 1]$ is the discount factor. Most c-MARL tasks can be regarded as a MMDP, and agents hope to learn a stationary policy $\pi(\cdot|s)$ to maximize the discounted return $G_t=\sum\limits_{k=0}^{\infty}{\gamma^k r_{t+k}}$.

When it comes to adversarial attacks, we introduce an adversary $\nu(\cdot) \in B(\cdot)$ to perturb the elements in MMDP, where $B(\cdot)$ is the available perturbation set. $(\cdot)$ can be state $s$, action $a$ or reward $r$. The goal of adversarial attack is to minimize the discounted return $G$ by perturbing elements in MMDP. Generally, due to the limited perturbation budget and attack feasibility, the adversaries are encouraged to perturb the elements in MDP as small as they could, while achieving attack as strong as they could. Thus, attackers have the full control of victim model and in most works, they choose to attack on one of the three elements which is suitable for their specific conditions.

\subsubsection{Attacks towards States} The goal of attacks towards states aims to perturb the state observed by agents, such that agents will perform erroneous actions that harm the final reward. The goal of attacks towards state can be formulated as below:
\begin{equation}
\begin{aligned}
  &\min_{\nu(s)} G_t=\sum\limits_{k=0}^{\infty}{\gamma^k r_{t+k}}, \\
  &\text{s.t.  } \nu(s)\in B(s), s'\sim P(s'|s,\{a^i\}), \{a^i\}\sim \pi(\cdot|\nu(s)),
  \label{eq:atk_state}
\end{aligned}
\end{equation}

Note that state-based attack is very similar to adversarial attacks in computer vision, or natural language processing, which perturb model input at test time to mislead the output at test time. Therefore, methods perturbing states usually apply gradient-based attack (\eg FGSM, PGD, \etc) to generate the adversarial perturbation in a white-box setting. For black-box settings, the transferability of adversarial examples renders it possible to train a surrogate model, generate white-box perturbations, then transfer them to the black-box policy. Attacks based on query is also feasible. 

\subsubsection{Attacks towards Rewards} Applied in training time, attacks towards rewards modify rewards given to agents to interfere the policy, such that the attacked agent cannot achieve its desired goal. Given adversary $\nu$ that perturbs the reward, the process can be formulated as:
\begin{equation}
\begin{aligned}
  &\min_{\nu(r)} G_t=\sum\limits_{k=0}^{\infty}{\gamma^k r_{t+k}}, \\
  &\text{s.t.  } \nu(r) \in B(r), s'\sim P(s'|s,\{a^i\}), \{a^i\}\sim \pi(\cdot|s,\nu(r)).
  \label{eq:atk_reward}
\end{aligned}
\end{equation}
where $\pi(a|s,\nu(r))$ means the policy $\pi$ is learned from perturbed reward $\nu(r)$. This type of attack usually do not need to have any knowledge of the model. Instead, it misleads the policy by flipping the sign of rewards at certain time. Previous works \cite{han2018reinforcement} have shown that only a certain number (\eg, 5\% of all experiences) of flipping operation can destroy the training process. This can be extremely harmful when the reward signal is corrupted or hijacked.

\subsubsection{Attacks towards Actions} Attacks towards actions directly perturb selected actions without modifying rewards or observations. Specifically, attacks towards actions substitute the action given by original policy to the action given by an adversarial policy, or add noise to the original policy to minimize the total goal. The formulation of attacks towards actions can be listed as follows:
\begin{equation}
\begin{aligned}
  &\min_\theta G_t=\sum\limits_{k=0}^{\infty}{\gamma^k r_{t+k}}, \\
  &\text{s.t.  } \nu(\pi) \in B(\pi), s'\sim P(s'|s,\{a^i\}), \{a^i\}\sim \nu(\pi(\cdot|s)).
  \label{eq:atk_action}
\end{aligned}
\end{equation}
There are several approaches about how to perturb the action. For continuous action space, adversary can add proper perturbations on the original action. Additionally, it is effective to train an adversarial policy $\pi_{adv}$ to replace the original policy $\pi$. Attacks towards actions require the control of model outputs, and accordingly, it can degrade the model performance by a large margin.

\subsection{Proposed Method}

Based on the formulation above, we propose our MARLSafe method with the help of three types of attacks. These attacks can (1) comprehensively cover the elements of MMDP and test the robustness from multiple aspects, (2) test the policy at training time and test time, and (3) can run in white-box and black-box settings. The overall framework of MARLSafe is given in Fig. \ref{fig:framework}.

\subsubsection{State Test} To test the robustness in the dimension of state, we apply gradient based attack to the observation of agents. In c-MARL settings, agents follow CTDE framework and communicate with each other only by their observations at test time. The perturbed observations mislead the attacked agent to deviate from the cooperation. To simplify the attack, we add perturbations by using fast gradient sign method (FGSM)\cite{goodfellow2014explaining}, where we execute untargeted attack with the objective of minimizing the probability of original optimal action. As a result, our adversarial perturbations reduce the output logits of optimal actions, and induce the policy to choose worse actions. Define $a^*=\arg \max_a \pi(s)$ as the optimal action selected by policy, and $Q(s, a^*;\pi)$ refers to the output logit of policy network $\pi$ for $s$ and $a^*$. The attack in state test can be formulated as below:

\begin{equation}
\begin{aligned}
  & \nu(s) = s - \epsilon\cdot\text{sign}(\nabla Q_s(s, a^*;\pi)).
  \label{eq:test_state}
\end{aligned}
\end{equation}

\subsubsection{Reward Test} Reward poisoning is a threatening attack method for RL policies at training time. In c-MARL settings, the training process is centralized, and the environment usually returns a total reward rather than a group of rewards. Motivated by \cite{han2018reinforcement}, we flip the sign of a certain percent $k\%$ of rewards during training time, poisoning rewards to prevent the model from acquiring a good policy. As the training data is collected one episode a time, we choose the max $k\%$ reward in an episode to flip their sign. When $k=0$, it becomes a normal training process, and when $k=100$, the policy will aim at minimizing the team reward. Define $r_{thresh}$ as the threshold at the $k\%$ rewards of all time steps, and we can formulate our testing method as below:

\begin{align}
  & \nu(r)=\left \{ 
  \begin{aligned}
  &r, & r \leq r_{thresh} \\
  &-r, & r > r_{thresh}
  \end{aligned}
  \right.
  \label{eq:test_reward}
\end{align}

\subsubsection{Action Test} In c-MARL settings, we use the powerful black-box attack method \emph{adversarial policy} \cite{gleave2019adversarial} as our testing method. However, vanilla adversarial policy is formulated as a zero-sum Markov game between two opposing agents, which is not suitable for c-MARL settings. For MARLSafe, we propose the testing method that we get the control of one agent as a "traitor", and train the "traitor" to maximally perturb other agents with their policies fixed. The reward of the adversarial policy can be set to the opposite number of original team reward (\ie, the traitor seeks to minimize the reward of collaborated agents with fixed policy). Define the reward of adversarial policy as $r'$, the formulation of action test is listed in \cref{eq:test_action}.

\begin{equation}
\begin{aligned}
  & \nu(\pi)=\pi_\alpha(s), \\
  & \text{ s.t. } \max_{\nu(\pi)}\sum\limits_{k=0}^\infty \gamma^k r'_{t+k}
  \label{eq:test_action}
\end{aligned}
\end{equation}

\section{Experiments}

In this section, we conduct experiments to demonstrate the effectiveness of MARLSafe. We choose \textit{StarCraftII Multi-Agent Challenge} (\textit{SMAC})\cite{samvelyan19smac} as our experimental environment. We use \textit{EPyMARL} framework as our testbed. All of our experiments are conducted on a server with 3 NVIDIA RTX 2080Ti GPUs and a 26-core 2.10GHz Intel(R) Xeon(R) Gold 6230R CPU.

\subsection{Experiment Settings}

\textbf{MARL Algorithm} We choose QMIX\cite{rashid2018qmix} and MAPPO\cite{yu2021surprising} as algorithms to test. QMIX and MAPPO are popular algorithms with CTDE framework in c-MARL tasks, thus deserving a test. QMIX is based on Q-learning, while MAPPO is based on policy gradient. In QMIX, agents share a deep Q-network for decentralized execution, and a Q-value mixer is applied for centralized training. In MAPPO, agents select actions via an actor network, and actions of all agents are evaluated by a critic network. The hyperparameters of MARL algorithms is consistent with \textit{EPyMARL}.

\textbf{SMAC Maps} We choose 2s3z (2 Stalkers and 3 Zealots) and 11m (11 Marines) as our experiment maps, where red team controlled by agents and blue team controlled by computer have same units. Note that the original \textit{SMAC} does not contain the map "11m". The 11m map is modified from the original map "10m\_vs\_11m", to balance the number of units for two players and give fair comparison in action test. The goal of normal agents in \textit{SMAC} maps is to kill enemy units as much as possible, and the game ends when one team lose all units or the time step reach the limit. In our setting of action test, the first agent will be controlled as a "traitor". In 2s3z, the "traitor" is a Stalker. We select the difficulty of computer as level 7 (hardest possible).

\textbf{Hyperparameters of Attack} In state test, we perform FGSM attack in $\ell_\infty$-norm and set $\epsilon=0.05$. In reward test, we set filp rate $k\%=10\%$. In action test, we fix other agent's policy and train the "traitor" agent with deep recursive Q-network (DRQN)\cite{hausknecht2015deep}. The traitor do not have access to global state or observation of other agents. As for the reward of traitor, it receives a positive reward when allies get damaged or die, and receives a negative reward when when enemies get damaged or die. Winning the game will receive negative rewards, and losing the game will receive positive rewards. The reward is normalized to $[-20, 20]$.

\textbf{Evaluation Metrics} The performance of a policy in \textit{SMAC} environment can be evaluated by metrics below: win rate (WR), team reward (TR), mean number of dead allies (mDA), and mean number of dead enemies (mDE). In our experiments, we calculate and show these 4 metrics to evaluate the robustness of MARL policies. We test 32 episodes of games for each experiment to calculate these metrics.

\begin{figure*}
  \centering
  \begin{subfigure}{0.32\linewidth}
    \includegraphics[width=\linewidth]{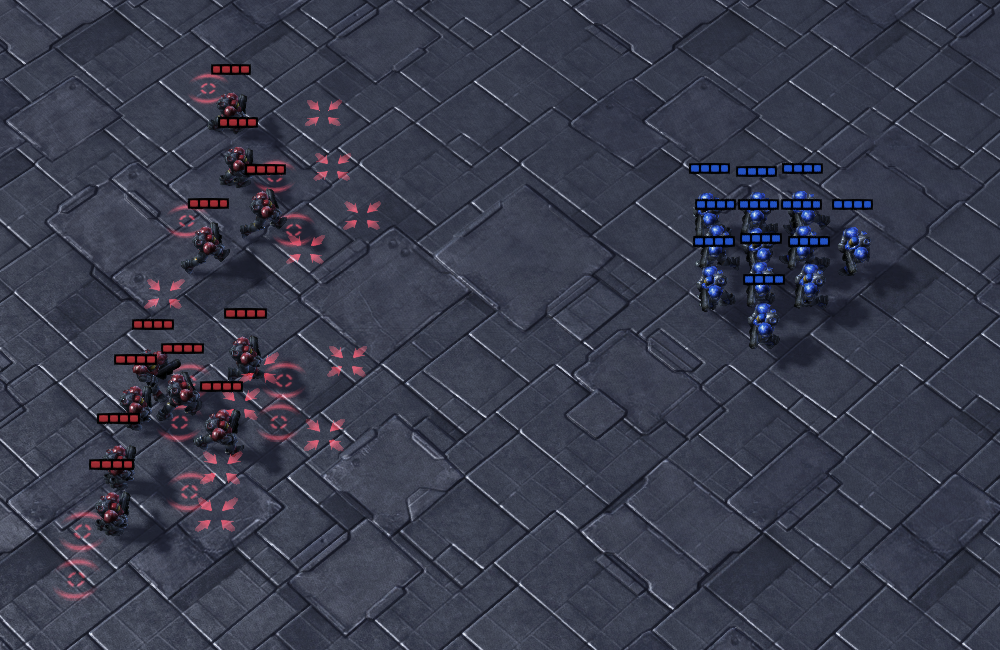}
    \caption{attack towards states}
    \label{fig:state}
  \end{subfigure}
  \hfill
  \begin{subfigure}{0.32\linewidth}
    \includegraphics[width=\linewidth]{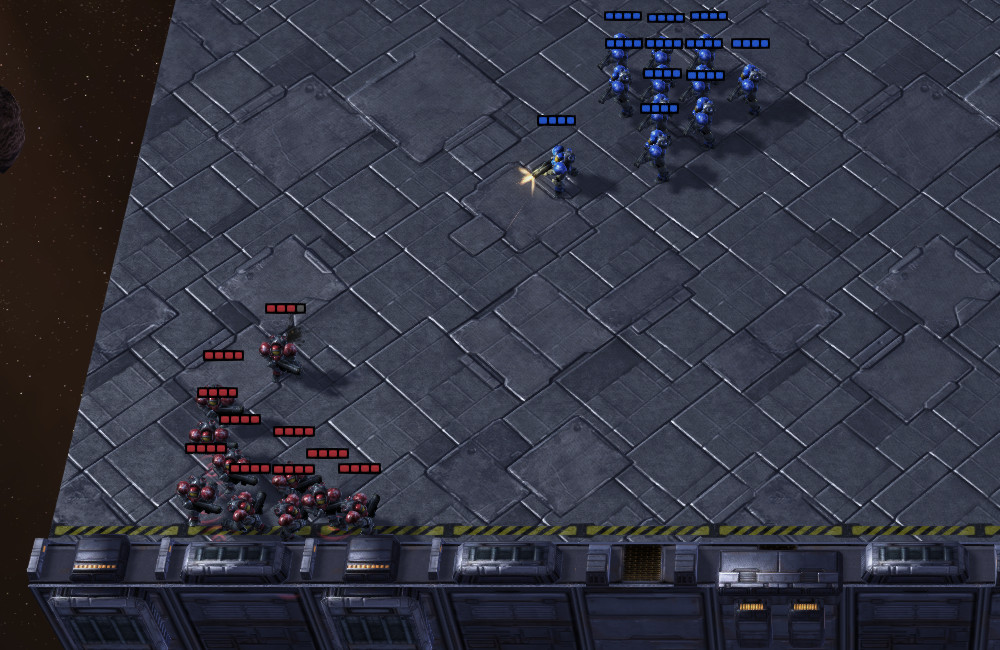}
    \caption{attack towards rewards}
    \label{fig:reward}
  \end{subfigure}
  \hfill
  \begin{subfigure}{0.32\linewidth}
    \includegraphics[width=\linewidth]{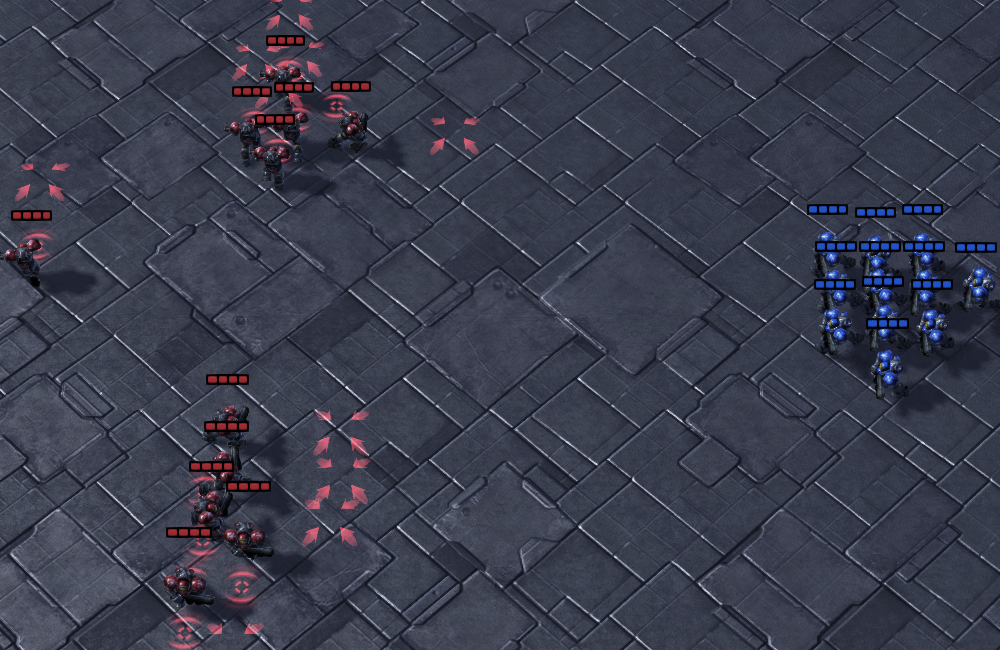}
    \caption{attack towards actions}
    \label{fig:action}
  \end{subfigure}
  \caption{Illustration of different behaviors of MARL agents under three types of attacks. Agents under state-based attack act relatively normal, but failed to cooperate. Agents under reward-based attack jointly flee from opponents. Traitor under action-based attack first flee away, then exert influence on normal agents.}
  \label{fig:behavior}
  \vspace{-0.1in}
\end{figure*}

\begin{figure}[t]
  \centering
  \begin{subfigure}{0.49\linewidth}
    \includegraphics[width=\linewidth]{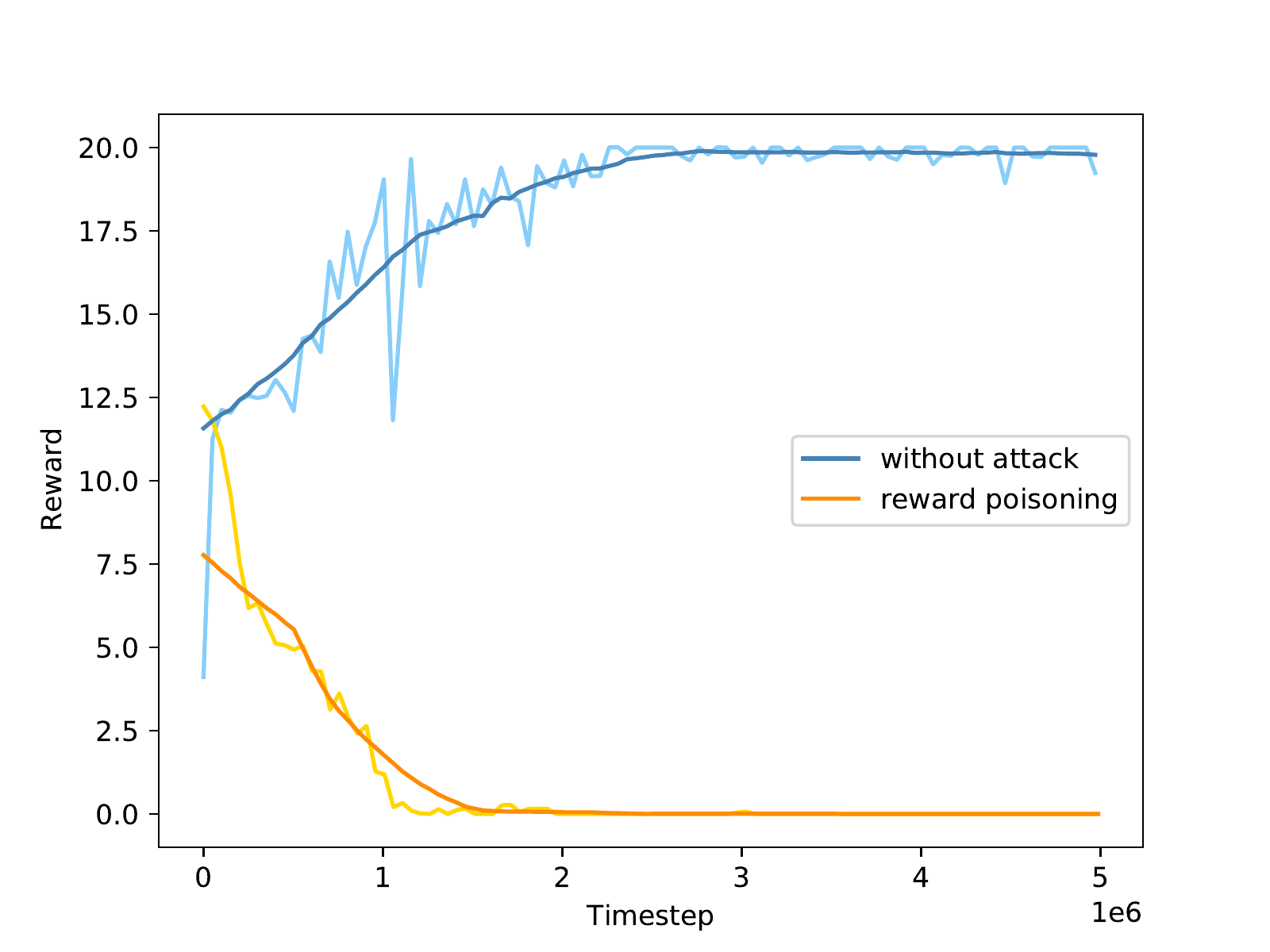}
    \caption{MAPPO, 2s3z}
    \label{fig:2s3z}
  \end{subfigure}
  \hfill
  \begin{subfigure}{0.49\linewidth}
    \includegraphics[width=\linewidth]{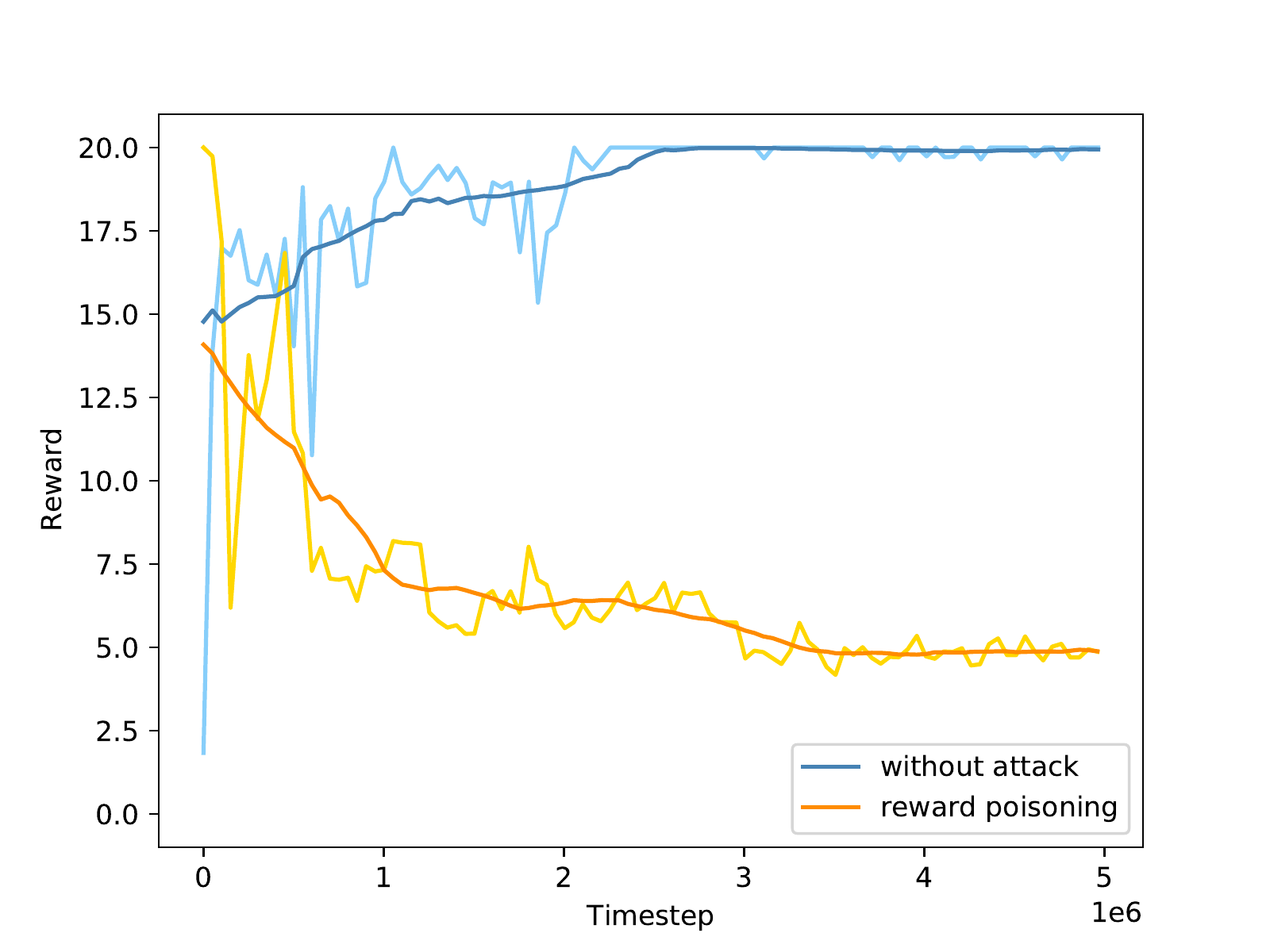}
    \caption{MAPPO, 11m}
    \label{fig:11m}
  \end{subfigure}
  \caption{Reward curve of normal agents and reward-attacked agents in an episode, during training.}
  \label{fig:rewards}
  \vspace{-0.3in}
\end{figure}

\subsection{Experimental Results}

\begin{table}
  \centering
  \begin{tabular}{@{}cccccc@{}}
    \toprule
    Map                   & Algorithm & WR & TR & mDA & mDE \\
    \midrule
    \multirow{2}{*}{2s3z} & QMIX      & 96.88\% & 19.74 & 0.84 & 4.94 \\
                          & MAPPO     & 100.00\% & 20.00 & 0.84 & 5.00 \\
    \hline
    \multirow{2}{*}{11m}  & QMIX      & 100.00\% & 20.00 & 2.00 & 11.00 \\
                          & MAPPO     & 100.00\% & 20.00 & 2.81 & 11.00 \\
    \bottomrule
  \end{tabular}
  \caption{Performance of QMIX and MAPPO without attack. Both algorithms were able to outperform hardest build-in AI of Starcraft with a very high win rate.}
  \label{tab:no_atk}
\end{table}

\textbf{Performance without Attack} The performance of QMIX and MAPPO without attack is listed in \cref{tab:no_atk}. Without attack, the performance of QMIX and MAPPO greatly surpasses the hardest build-in AI inside the StarCraftII game. Both in 2s3z and 11m, the winning rate reaches 100\% with a maximum reward. The great performance proves the effectiveness of these MARL algorithms.

\begin{table}
  \centering
  \begin{tabular}{@{}cccccc@{}}
    \toprule
    Map                   & Algorithm & WR & TR & mDA & mDE \\
    \midrule
    \multirow{2}{*}{2s3z} & QMIX      & 9.38\% & 11.85 & 4.72 & 1.69 \\
                          & MAPPO     & 65.62\% & 17.91 & 3.34 & 4.25 \\
    \hline
    \multirow{2}{*}{11m}  & QMIX      & 0.00\% & 9.76 & 10.94 & 5.53 \\
                          & MAPPO     & 31.25\% & 14.62 & 9.69 & 8.66 \\
    \bottomrule
  \end{tabular}
  \caption{Performance of QMIX and MAPPO under state test. Both algorithms shows weak robustness, while MAPPO is relatively robust.}
  \label{tab:state}
\end{table}

\begin{table}
  \centering
  \begin{tabular}{@{}cccccc@{}}
    \toprule
    Map                   & Algorithm & WR & TR & mDA & mDE \\
    \midrule
    \multirow{2}{*}{2s3z} & QMIX      & 0.00\% & 5.32 & 4.41 & 0.06 \\
                          & MAPPO     & 0.00\% & 0.00 & 3.72 & 0.00 \\
    \hline
    \multirow{2}{*}{11m}  & QMIX      & 0.00\% & 0.00 & 2.09 & 0.00 \\
                          & MAPPO     & 0.00\% & 5.89 & 11.00 & 0.16 \\
    \bottomrule
  \end{tabular}
  \caption{Performance of QMIX and MAPPO under reward test. None of the algorithms are robust under reward-based attack.}
  \label{tab:reward}
\end{table}

\begin{table}
  \centering
  \begin{tabular}{@{}cccccc@{}}
    \toprule
    Map                   & Algorithm & WR & TR & mDA & mDE \\
    \midrule
    \multirow{2}{*}{2s3z} & QMIX      & 0.00\% & 10.07 & 4.97 & 1.34 \\
                          & MAPPO     & 0.00\% & 11.59 & 5.00 & 2.13 \\
    \hline
    \multirow{2}{*}{11m}  & QMIX      & 6.25\% & 9.88 & 10.69 & 4.81 \\
                          & MAPPO     & 0.00\% & 10.53 & 11.00 & 6.31 \\
    \bottomrule
  \end{tabular}
  \caption{Performance of QMIX and MAPPO under action test. None of the algorithms are robust under action-based attack.}
  \label{tab:action}
\end{table}

\textbf{Performance under State Test} We apply our state test on QMIX and MAPPO, whose results are showed in \cref{tab:state}. The results show the vulnerability of these two algorithms, drastically reducing the winning rate of QMIX and MAPPO from 100\% to lower than 15\%. State-based attacks came as a "natural" attack. The replay demonstrates that all agents behaves as they are on battle with their opponent naturally, but ends up losing the game. When comparing two algorithms, we find that agents trained with MAPPO algorithm is more robust in the dimension of state than those trained with QMIX. Considering that the decentralized agent network structures of QMIX and MAPPO are identical, we hypothesis the robustness came from the training process, and the centralized training network (\ie mixer network or critic network) might play an important role in model robustness.

\textbf{Performance under Reward Test} The results of reward test are listed in \cref{tab:reward}. We notice that win rates in these experiments are equally 0\%. However, the mean number of dead allies does not reach the number of all allies. According to the detailed results, not all allies are killed when the time step reaches the limit, and the game ends with nobody winning. The trained policy behaves as all agents fleeing from their enemies. They spare no effort to fight against enemies or quickly surrender and get killed. It possibly results from the characteristic of reward attack, which only flips the sign of the maximum part of rewards. When enemies get damaged or killed, the reward increases rapidly and thus is likely to be perturbed, thus avoided. When agents died, the reward decreases, and is unlikely to get perturbed, As a result, agents get punished by being killed and develop the behaviour of avoid getting killed, but not killing opponents. \cref{fig:rewards} shows the total rewards by time step during training. Despite perturbing only 10\% of the total reward, the algorithm learns a corrupted policy, and the real reward continues to go down. Due to the high effectiveness of reward test, we conclude that robustness at the dimension of reward is usually overlooked yet vulnerable. It is necessary to pay more attention to reward poisoning attacks.

\textbf{Performance under Action Test} \cref{tab:action} shows the performance of QMIX and MAPPO under action test. We can draw a conclusion that the "traitor" controlled by adversarial policy leads to a great failure of battles. Interestingly, in the original \textit{SMAC} map "10m\_vs\_11m", 10 Marines controlled by MARL policies can easily defeat 11 Marines controlled by the computer with almost 100\% win rate. However, when the traitor was added, the agents performed much worse with stronger ally numbers, stongly proofing the vulnerability of MARL policies. On the other hand, agents controlled by QMIX can sometimes defeat the computer in 11m map, which implies its better robustness towards action test.

\textbf{Different Behaviors of Agents under Attacks} \cref{fig:behavior} presents the behavior of MARL agents under different types of attacks. We can clearly see the various behavior of agents and infer the characteristic of these attacks. Under the attack toward states, agents seem to try to behave as normal agents, but their actions become chaotic and non-cooperative. Clearly, attack toward states only suppresses the probability of the optimal action, so agents tend to choose suboptimal actions, which are sometimes effective but cannot achieve 100\% win rate since not being optimal. Under the attack towards rewards, agents seem to flee from their enemies. As the sign of greater reward flipped, agents avoid causing severe damages to their enemies, but they also avoid death because they cannot get any reward after they die. These limitations result in the fleeing behavior of agents. Under the attack towards actions, the adversarial policy hide behind the team while others attack as normal. However, the action of the adversarial policy affects the decision of its teammates, leading them to be defeated. After all normal agents die, the adversarial agent moves forward and quickly get killed. The adversarial policy learned by reinforcement learning acquire the optimal action to lose the game.

\section{Conclusion}

In this paper, we propose MARLSafe, a robustness testing framework for c-MARL algorithms. First, we formulate the existing attack method in the formulation of MDP, and categorize them by MDP elements: state, reward and action. Moreover, we propose a robustness testing method from multi aspects, and propose several adversarial attack method c-MARL settings to test the robustness of c-MARL algorithms. To the best of our knowledge, MARLSafe is the first paper to test the robustness of c-MARL algorithms from multiple aspects, and our method could attack state-of-the-art c-MARL algorithm with high performance degradation. The results of our experiments indicate that c-MARL algorithms are facing severe robustness problems, and it is necessary to explore comprehensive defense methods that jointly covers the aspect of state, action and reward.


{\small
\bibliographystyle{plain}
\bibliography{egbib}
}

\end{document}